\documentclass[aoas,preprint]{imsart}

\RequirePackage{amsthm,amsmath,amsfonts,amssymb}
\RequirePackage[authoryear]{natbib}
\RequirePackage{graphicx}
\RequirePackage{mathtools}
\startlocaldefs

\endlocaldefs

\begin{document}

\begin{frontmatter}
\title{Bayesian Multivariate Sparse Functional Principal Components Analysis with Application to Longitudinal Microbiome Multi-Omics Data}
\runtitle{multivariate SFPCA for Longitudinal Microbiome Multi-Omics Data}

\begin{aug}
\author[A]{\fnms{Lingjing} \snm{Jiang}\ead[label=e1,mark]{lij014@health.ucsd.edu}},
\author[B]{\fnms{Chris} \snm{Elord}\ead[label=e2]{chris.elrod@juliacomputing.com}},
\author[C]{\fnms{Jane} \snm{J. Kim}\ead[label=e6]{janekim@health.ucsd.edu}},
\author[D]{\fnms{Austin}
\snm{D. Swafford}\ead[label=e3]{adswafford@eng.ucsd.edu}},
\author[C,D,E,F]{\fnms{Rob} \snm{Knight}\ead[label=e4]{robknight@health.ucsd.edu}}
\and
\author[A]{\fnms{Wesley} \snm{K. Thompson}\ead[label=e5, mark]{wkthompson@health.ucsd.edu}}
\address[A]{Herbert Wertheim School of Public Health and Human Longevity Science, 
University of California San Diego, 
\printead{e1,e5}}

\address[B]{Julia Computing, 
\printead{e2}}

\address[C]{Department of Pediatrics,
University of California San Diego, 
\printead{e6,e4}}

\address[D]{Center for Microbiome Innovation,
University of California San Diego,
\printead{e3,e4}}

\address[E]{Department of Computer Science and Engineering,
University of California San Diego, 
\printead{e4}}

\address[F]{Department of Bioengineering,
University of California San Diego,
 \printead{e4}}

\end{aug}

\begin{abstract} 
Microbiome researchers often need to model the temporal dynamics of multiple complex, nonlinear outcome trajectories simultaneously. This motivates our development of {\it multivariate Sparse Functional Principal Components Analysis} (mSFPCA), extending existing SFPCA methods to simultaneously characterize multiple temporal trajectories and their inter-relationships. As with existing SFPCA methods, the mSFPCA algorithm characterizes each trajectory as a smooth mean plus a weighted combination of the smooth major modes of variation about the mean, where the weights are given by the component scores for each subject. Unlike existing SFPCA methods, the mSFPCA algorithm allows estimation of multiple trajectories simultaneously, such that the component scores, which are constrained to be independent within a particular outcome for identifiability, may be arbitrarily correlated with component scores for other outcomes. A Cholesky decomposition is used to estimate the component score covariance matrix efficiently and guarantee positive semi-definiteness given these constraints.  Mutual information is used to assess the strength of marginal and conditional temporal associations across outcome trajectories. Importantly, we implement mSFPCA as a Bayesian algorithm using \textsf{R} and \textsf{stan}, enabling easy use of packages such as PSIS-LOO for model selection and graphical posterior predictive checks to assess the validity of mSFPCA models. Although we focus on application of mSFPCA to microbiome data in this paper, the mSFPCA model is of general utility and can be used in a wide range of real-world applications.

\end{abstract}


\end{frontmatter}

\section{Introduction}
Numerous disorders, including heritable immune-mediated diseases such as inflammatory bowel disease (IBD) and asthma, neurological conditions including autism, and genetically driven diseases such as cancer, have been linked to dysregulation of human microbiota~\citep{holleran2018fecal, lloyd2019multi, frati2019role, sharon2019human, ballen2016infection}. However, the complex influences of microbiota on human health are not yet functionally understood. To understand the links between the human microbiome and disease, it is necessary to determine which microbe genes are being expressed and the timing of their expression~\citep{sberro2019large}. Thus, in addition to obtaining microbiome data using 16S ribosomal RNA gene sequencing or whole genome shotgun sequencing~\citep{kuczynski2010direct, ranjan2016analysis, gill2006metagenomic}, an increasing number of studies are also collecting transcriptomics data to understand microbial gene expression, proteomics data to study expressed proteins, and metabolomics data to define the functional status of host-microbial relationships~\citep{integrative2014integrative, lloyd2019multi, bouslimani2019impact}. This complex combination of data types, called {\it microbiome multi-omics}, is essential for understanding the links between microbial communities and disease and may enable translation of microbiome research into effective treatments.

An increasing number of microbiome multi-omics studies are longitudinal, aimed at simultaneously characterizing microbiome and host temporal changes to provide a more comprehensive picture of dynamics during healthy and diseased states~\citep{integrative2014integrative, lloyd2019multi, vatanen2018human, stewart2018temporal}. Despite these breakthroughs in microbiome study designs and data collection, few statistical methods are available to analyze these complex longitudinal omics data. Recently, several new methods based on network analysis were developed for multi-omics integration of microbiome data in cross-sectional studies~\citep{jiang2019microbiome, morton2019learning}, however, analytical methods for longitudinal microbiome multi-omics data are still in their infancy. The challenges include irregular timing and frequency across subjects, unmatched time points between different data types, non-linear temporal patterns, missing data, and high individual variability~\citep{bodein2019generic}. 

The statistical framework of functional data analysis (FDA) was introduced by Ramsay and Silverman~\citep{ramsay1997functional}, wherein the basic unit of information is the entire function, such as a curve or an image. Functional principal component analysis (FPCA) has been a widely used tool in FDA. The fundamental aim of FPCA is to reduce dimensionality by capturing the principal modes of smooth variation. FPCA summarizes subject-specific deviations from the mean curve via the coordinates ({\it principal component scores}) of the basis spanned by the principal components~\citep{di2009multilevel}. Existing FPCA methods include smoothed FPCA approaches based on roughness penalties~\citep{rice1991estimating}, extensions to sparsely sampled functional data ~\citep{james2000principal, yao2005functional, peng2009geometric, di2014multilevel, kidzinski2018longitudinal}, and the asymptotic properties of FPCA~\citep{hall2006properties, li2010uniform}. Most FPCA methods development has focused on univariate functional data. ~\citet{chiou2014multivariate} proposed a multivariate FPCA method to simultaneously model multiple temporal outcomes and infer the component dependencies through pairwise cross-covariance functions. However, this method is limited to functional data as classically conceived, where the curves are observed longitudinally over densely and consistently sampled time points. 

To meet the need for modeling irregularly and sparsely sampled, non-linear  multivariate microbiome multi-omics trajectories, we developed multivariate sparse functional principal components analysis (mSFPCA). The major novelty of our approach is that it focuses on a set of functions which may covary in complex ways. Smoothing is accomplished by retaining a low-dimensional set of  PC functions, as in~\citet{james2000principal}. The PC scores across outcomes are modeled jointly via a constrained covariance matrix, efficiently estimated by Cholesky decomposition. Our proposed method allows for simultaneously characterizing multiple temporal measurements, such as microbiome, metabolome, inflammatory markers, and  self-report measures, and to infer the temporal associations among these measures both marginally and conditionally, based on estimation of marginal and partial mutual information. The model is implemented using a Bayesian formulation, instantiated with Hamiltonian Markov Chain Monte Carlo (MCMC) methods in \textsf{stan} to sample from the posterior distribution of the model parameters. Our Bayesian implementation enables the usage of PSIS-LOO for model selection and graphical posterior predictive checks to assess the validity of mSFPCA models. While we focus on application of mSFPCA to microbiome data in this paper, the mSFPCA model is of general utility and can be used in many applications.

The remainder of the paper is organized as follows. Section 2 reviews sparse functional principal component analysis (SFPCA), and introduces multivariate SFPCA, our statistical framework for longitudinal microbiome multi-omics data. Section 3 describes extensive simulation studies to evaluate the performance of mSFPCA in realistic settings. Section 4 describes the application of our methodology to a challenging longitudinal microbiome multi-omics data on type 2 diabetes. Section 5 presents our conclusion. To ensure reproducibility of our results, accompanying software, simulations and analysis results are posted at https://github.com/knightlab-analyses/mfpca-analyses.

\section{Methodology}
\subsection{Sparse functional principal components analysis}
The classical assumption of functional data analysis is that each trajectory is sampled over a dense grid of time points common to all individuals \citep{ramsay2007applied}. However, in practice, trajectories are often measured at an irregular and sparse set of time points that can differ widely across individuals. To address this issue, \citet{james2000principal} proposed {\it sparse functional principal components analysis} (SFPCA) using a reduced rank mixed-effects framework. Let $Y_i (t)$ be the measurement at time $t$ for the $i$th individual, $\mu(t)$ the overall mean function, $f_j$ the $j$th principal component function and ${f}= [(f_1,f_2,…,f_k)]^T$, where $k$ is the number of principal components. Then the \citet{james2000principal} SFPCA model is given by 
\begin{equation}
    Y_i(t) = \mu(t) + \sum_{j=1}^{k}f_j(t) \alpha_{ij} + \epsilon_i(t), \quad i = 1, ..., N
\end{equation}
subject to the orthogonality constraint $\int f_j  f_l = \delta_{jl}$, the Kronecker $\delta$. The vector 
$\alpha_i = (\alpha_{i1},\ldots,\alpha_{ik})^T$ is the component weights for the $i$th individual and $\epsilon_i (t)$ is a normally-distributed residual, independent across subjects and across times within subject. The functions $\mu$ and ${f}$ are approximated using cubic splines to allow a smooth but flexible fit. Let $b(t)$ be a cubic spline basis with dimension $q > k$.
The spline basis is orthonormalized so that $\int b_j b_l = \delta_{jl}$.
Let $\Theta$ and $\theta_\mu$ be, respectively, a $q \times k$ matrix and a $q$-dimensional vector of real-valued coefficients. For each individual $i$, denote their measurement times by $t = (t_{i1},t_{i2},…,t_{in_i})^T$, and let  $Y_i= (Y_i (t_{i1}),…,Y_i (t_{in_i }))^T$ be the corresponding real-valued observations. Then $B_i= ( b(t_{i1}),…, b(t_{in_i}))^T$ is the $n_i \times q$ spline basis matrix for the $i$th individual. The reduced rank model can then be written as 
\begin{align}
    Y_i = B_i \theta_\mu + B_i \Theta \alpha_i + \epsilon_i, \quad i = 1, ..., N, \label{eq:fpca} \\
    \Theta^T \Theta = I, \quad \alpha_i \sim N(0, D), \quad \epsilon_i \sim N(0, \sigma_\epsilon^2 I_{n_i}), \nonumber 
\end{align} 
where the covariance matrix $D$ is restricted to be diagonal and $I_{n_i}$ is the $n_i \times n_i$ identity matrix. Various fitting approaches, such as the EM algorithm, kernel smoothing and Newton-Raphson algorithm, have been proposed to estimate parameters of the SFPCA model \citep{james2000principal, yao2005functional, peng2009geometric}. These approaches then use model selection techniques, such as cross-validation, Akaike information criterion (AIC) and leave-one-curve-out cross-validation, to select the dimension of the basis and the number of principal components. However, due to their reliance on assumptions such as normally-distributed component scores and residuals, these models need to be carefully examined when applied to real data \citep{kidzinski2018longitudinal}. 

\subsection{Bayesian SFPCA}
\citet{jiang2020bayestime} proposed an SFPCA model in a Bayesian framework to allow for flexible prior specification and implementation of model selection and assessment methods. This Bayesian implementation used the Hamiltonian MCMC sampling algorithm in \textsf{Stan} \citep{carpenter2017stan}. The real-valued observations $Y_i(t)$ are first standardized to have mean zero and standard deviation one. The prior distributions for parameters in Eq. (\ref{eq:fpca}) were chosen as follows:

\begin{align*}
    \theta_\mu  &\sim N_q (0,I_q ) \\
    \alpha_i &\sim N_k (0,I_k ) \\
    \Theta_j &\sim N_q (0,I_q ), j=1,\ldots,k \\
    \epsilon_i  &\sim N_{v_i} (0,\sigma_\epsilon^2 I_{v_i}) \\
    \sigma_\epsilon &\sim Cauchy(0,1),
\end{align*}
where $\Theta_j$ is the $j$th column of the loading matrix $\Theta$, and $v_i$ is the total number of visits for the $i$th subject. This Bayesian implementation enables use of 
leave-one-out cross-validation with Pareto-smoothed important sampling (PSIS-LOO) \citep{vehtari2017practical} to perform model selection on the number of principal components $k$ 
and the number of basis functions $q$.  Moreover, model fit can be assessed via diagnostic plots from PSIS-LOO as well as the graphical posterior predictive checks obtained from simulating posterior predictive data \citep{gelman1996posterior, gabry2019visualization}. This Bayesian implementation thus offers a flexible and comprehensive solution to real-data SFPCA applications.

\subsection{Multivariate SFPCA}
Here, we extend our previous Bayesian SFPCA to simultaneously model multiple trajectories, and to quantify their marginal and conditional temporal associations. Let $\boldsymbol{Y_i}(t)$ denote the $P$-dimensional observed response at time $t$ for subject $i$, which can be modeled by the multivariate Functional PCA (mFPCA) model as
\begin{equation}
\boldsymbol{Y_i}(t) = \boldsymbol{\mu_i}(t) + \boldsymbol{f}(t)^T \boldsymbol{\alpha_{i}} + \boldsymbol{\epsilon_i}(t), i = 1, ..., N,
\end{equation}
where $\boldsymbol{\mu_i} = (\mu_{i,1}(t), ..., \mu_{i,P}(t))^T$ is the overall mean response of $P$ trajectories for subject $i$, $\boldsymbol{f}(t)^T = diag(f_{1}(t)^T, ..., f_{P}(t)^T)$, with $f_{p}(t)$ the FPC functions corresponding to the $p$th trajectory at time $t$, $\boldsymbol{\alpha_{i}} = (\alpha_{i,1}, ..., \alpha_{i,P})$ is the vector of FPC scores for subject $i$, and $\boldsymbol{\epsilon_i}$ is the corresponding residuals.

In order to fit this model when the data are sampled at only a finite number of time points, we chose to fit $\boldsymbol{\mu_i}$ and $\boldsymbol{f}(t)$ using a basis of spline functions $\boldsymbol{B}$. Let $K_p$ be the number of FPCs, $Q_p$ be the corresponding number of basis functions, $V_{ip}$ be the total number of assessments for $i$th subject in the $p$th temporal measurement, $B_{ip}$ be the transpose of the cubic spline basis, and $\Theta_p$ be the corresponding FPC loadings. Then the total number of principal components across $p$ measurements are $K= \sum_{p=1}^P K_p$, the total number of basis functions are $Q= \sum_{p=1}^PQ_p$, and the total number of assessments for subject $i$ is $V_i= \sum_{p=1}^PV_{ip}$. The model for $\boldsymbol{Y_i}(t)$ can be written into the multivariate Sparse Functional PCA (mSFPCA) model as

\begin{equation}
\boldsymbol{Y_i} = \boldsymbol{B_i \theta_\mu} + \boldsymbol{B_i \Theta \alpha_i} + \boldsymbol{\epsilon_i}, i = 1, ..., N,
\label{eq:mSFPCA}
\end{equation}
where $\boldsymbol{Y_i}$ is a $P$-dimensional observed response,  residuals $\boldsymbol\epsilon_i \sim N_{V_i} (0, \sigma_{\boldsymbol\epsilon}^2 I_{V_i})$, 
$\boldsymbol{B_i}$ is a $V_i \times Q$ matrix with
\begin{equation*}
\boldsymbol{B_i} = 
\begin{bmatrix}
B_{i1} & 0^T & \cdots  & 0^T \\
0 & B_{i2} & \cdots  & 0^T \\
\vdots & \vdots & \ddots & \vdots \\
0 & 0 & \cdots & B_{iP}
\end{bmatrix},
\end{equation*}
where $B_{ip}$ is the $V_{ip} \times Q_p$ matrix of spline bases evaluated at assessment times ${\bf t}_{ip} = (t_{ip1},\ldots, t_{ipV_{ip}})$.
The $Q \times K$ matrix $\boldsymbol{\Theta}$ of FPC loadings
\begin{equation*}
\boldsymbol{\Theta} = 
\begin{bmatrix}
\Theta_1 & 0^T & \cdots & 0^T \\
0 & \Theta_2 & \cdots  & 0^T \\
\vdots & \vdots & \ddots & \vdots \\
0 & 0 & \cdots & \Theta_P
\end{bmatrix},
\end{equation*}
is subject to the orthonormality constraint $\boldsymbol \Theta^T \boldsymbol \Theta = I$.
The $K$-dimensional vector of FPC scores $\boldsymbol{\alpha_i} \sim N(0, \Sigma_\alpha)$, with $ \Sigma_\alpha$ is restricted to the form 
\begin{equation*}
\begin{bmatrix}
D_1 & C_{21}^T & \cdots & C_{P1}^T \\
C_{21} & D_2 & \cdots  & C_{P2}^T \\
\vdots & \vdots & \ddots & \vdots \\
C_{P1} & C_{P2} & \cdots & D_P
\end{bmatrix},
\end{equation*}
where $D_p$ is the within-trajectory diagonal covariance matrix (necessary for identifiability of within-trajectory FPCs) for the $p$th trajectory , and $C_{lm}$ is the covariance matrix for the $l$th and $m$th trajectories.
$ \Sigma_\alpha$ can be written as $ \Sigma_\alpha= S_\alpha R_\alpha S_\alpha$, where $S_\alpha$ is the diagonal matrix of standard deviations for the FPC scores,  and $R_\alpha$ is the correlation matrix restricted to the form
\begin{equation*}
\begin{bmatrix}
I_1 & R_{21}^T & \cdots & R_{P1}^T \\
R_{21} & I_2 & \cdots  & R_{P2}^T \\
\vdots & \vdots & \ddots & \vdots \\
R_{P1} & R_{P2} & \cdots & I_P
\end{bmatrix},
\end{equation*}
where $I_p$ is the $Q_p \times Q_p$ identity matrix corresponding to the $p$th trajectory. 

We implemented the sMFPCA model using the Hamiltonian Markov Chain Monte Carlo (MCMC) sampling algorithm in \textsf{Stan} to estimate parameters. The prior distributions for $\boldsymbol{\theta_\mu}$, $\boldsymbol{\Theta}$ and $\boldsymbol{\epsilon_i}$ are set as follows
\begin{equation}
\begin{aligned}
\boldsymbol{\theta_\mu} &\sim N_Q(0, I_Q) \\
\boldsymbol{\Theta_{kp}} &\sim N_{Q_p}(0, I_{Q_p}), \\
\boldsymbol{\epsilon_i} &\sim N_{V_i}(0,  \sigma_{\boldsymbol\epsilon}^2 I_{V_i}) \\
\sigma_{\boldsymbol\epsilon} &\sim Cauchy(0,1), \nonumber 
\end{aligned}
\end{equation}
where $\boldsymbol{\Theta_{kp}}$ is the $k$th column of the FPC loadings in the $p$th block. Leveraging this Bayesian implementation in \textsf{Stan}, we utilized PSIS-LOO for model selection, and diagnostics plots from PSIS-LOO and graphical posterior predictive checks for model diagnostics. 

\subsubsection{Orthonormality constraint}
A difficulty in implementing the Bayesian mSFPCA model is that the principal component loadings $\boldsymbol\Theta$ are not uniquely specified. 
For a given $K \times K$ rotation matrix $P$, if $\boldsymbol\Theta^* = \boldsymbol\Theta P$ and $\boldsymbol\Theta$ obeys the constraints in Eq.(\ref{eq:mSFPCA}), then $\boldsymbol\Theta^{*T} \boldsymbol\Theta^*= P^T \boldsymbol\Theta^T \boldsymbol\Theta P = I$, and hence $\boldsymbol\Theta$ is unidentifiable without additional restrictions. Instead of directly enforcing orthonormality when sampling from the conditional posteriors in the Bayesian model fitting, we sampled the parameters with no constraint on $\boldsymbol\Theta$ and then performed a {\it post hoc} rotation for each 
iteration of the MCMC algorithm to meet the orthonormality constraint.
Since the symmetric matrix $\boldsymbol\Theta \Sigma_\alpha \boldsymbol\Theta^T $ is identifiable and non-negative definite, we applied an eigenvalue decomposition $\boldsymbol\Theta \Sigma_\alpha \boldsymbol\Theta^T = V S V^T$, where $V$ is the $Q \times Q$ matrix of orthonormal eigenvectors, and $S$ is the diagonal matrix of eigenvalues, with the $Q$ positive eigenvalues ordered from largest to smallest. Let $\Theta^* = V_k$ denote the $Q \times K$ matrix consisting of the first $K$ eigenvectors of $V$, which satisfies $\boldsymbol\Theta^{*T} \boldsymbol\Theta^{*} = I$. 
Finally, we rotated $\Sigma_\alpha$ and FPC scores $\boldsymbol\alpha_i$, to obtain $\Sigma_\alpha^{*} = \boldsymbol\Theta^{T*} \boldsymbol\Theta \Sigma_\alpha \boldsymbol\Theta^T \boldsymbol\Theta^{*}$, and $\boldsymbol \alpha_i^* = \boldsymbol\Theta^{*T} \boldsymbol\Theta \boldsymbol\alpha_i$, so that $\boldsymbol\Theta^{*} \Sigma_\alpha^{*} \boldsymbol\Theta^{T*} = \boldsymbol\Theta \Sigma_\alpha \boldsymbol\Theta^T$, and $\boldsymbol\Theta^* \boldsymbol\alpha_i^{*} = \boldsymbol\Theta \boldsymbol\alpha_i$.

\subsubsection{Modeling covariance}
The covariance matrix of FPC scores $\Sigma_\alpha$ must be positive semi-definite and is restricted to be diagonal within-trajectory, it is a challenge to model this covariance matrix effectively. \citet{barnard2000modeling} proposed a separation strategy for modeling $\Sigma=SRS$ by assuming independent priors for the standard deviations $S$ and the correlation matrix $R$. To account for the dependent structure of correlations among different subsets of variables, \citet{liechty2004bayesian} proposed the common correlation model for $R$, which assumes a common normal prior for all correlations with the additional restriction that the correlation matrix is positive definite. However, the awkward manner in which $r_{ij}$, the ${ij}$th element in the correlation matrix $R$, is embedded in the full conditional posterior density, leading to use of a Metropolis-Hastings algorithm to update one coefficient $r_{ij}$ at a time \citep{liechty2004bayesian}. This consecutive updating procedure for correlation estimation is inefficient, and could lead to heavy computational cost when the correlation matrix is large or when the correlation has to be estimated separately from other parameters in mSFPCA model when implemented in \textsf{Stan} \citep{carpenter2017stan}. For example, in a simulated data with 3 temporal measurements from 100 subjects over 10 time points, it would take 40 hours for a mSFPCA model using Liechty's covariance estimation method to estimate all the parameters when implemented in Stan. However, the computational time can be reduced over 130 times (to only 18 minutes) by using our proposed method due to the avoidance of additional Metropolis-Hastings algorithm. 

To pursue an efficient numerical solution to the covariance estimation, we took advantage of the Cholesky decomposition \citep{nash1990cholesky} and imposed the diagonal constraint on the within-trajectory covariance matrices. Since the covariance matrix of FPC scores $\Sigma_\alpha$ has full rank with probability one, it has a unique Cholesky decomposition in the form of
\begin{equation}
\Sigma_\alpha = L L^T, \nonumber 
\end{equation}
where $L$ is a real lower triangular matrix with positive diagonal entries \citep{gentle2012numerical} . Given a lower triangular matrix $L$ divided into $P$ blocks, we have
\begin{equation*}
L = 
\begin{bmatrix}
L_{1,1} & 0 & \cdots & 0 \\
L_{2,1} & L_{2,2} & \cdots  & 0 \\
\vdots & \vdots & \ddots & \vdots \\
L_{P,1} & L_{P,2} & \cdots & L_{P,P}
\end{bmatrix},
\end{equation*}

\begin{equation*}
\text{then } L L^T = 
\begin{bmatrix}
L_{1,1}L_{1,1}^T & * & \cdots & * \\
L_{2,1}L_{1,1}^T & L_{2,1}L_{2,1}^T + L_{2,2}L_{2,2}^T & \cdots  & * \\
\vdots & \vdots & \ddots & \vdots \\
L_{P,1}L_{1,1}^T & L_{P,1}L_{2,1}^T +  L_{P,2}L_{2,2}^T& \cdots & L_{P,1}L_{P,1}^T + ... + L_{P,P}L_{P,P}^T
\end{bmatrix},
\end{equation*}
where $*$ denotes the transpose of the corresponding sub-diagonal block. To ensure that $L L^T$ is positive definite with diagonal within-block covariance matrices, the lower triangular Cholesky factor $L$ needs to meet the following two conditions:
\begin{enumerate}
\item Within-block covariance matrices $\sum_{m=1}^{M} L_{M,m} L_{M,m}^T, M = 1, ..., P$, are diagonal.
\item The diagonal entries of $L_{M,M}, M = 1, ..., P$ are positive.
\end{enumerate}
We will focus on defining the diagonal blocks $L_{M, M}$ to achieve these, and leave the off-diagonal blocks $L_{M, m}, m = 1, ..., M-1$ to be arbitrary, unconstrained (i.e. the unconstrained parameter elements from the Hamiltonian MCMC sampling). 

Let $D_M, M = 1, ..., P$ be the $M$th within-block covariance matrix, then 
\begin{equation}
\begin{aligned}
D_M &= \sum_{m=1}^{M} L_{M,m} L_{M,m}^T \\
&= L_{M,M} L_{M,M}^T + \sum_{m=1}^{M-1} L_{M,m} L_{M,m}^T, \\
L_{M,M} L_{M,M}^T &= D_M - \sum_{m=1}^{M-1} L_{M,m} L_{M,m}^T = A.
\end{aligned}
\end{equation}
Since all the off-diagonal elements of $D_M$ are known to be zero and the off-diagonal blocks $L_{M, m}, m = 1, ..., M-1$ are defined earlier with unconstrained estimates, we have thus defined all the off-diagonals of this matrix A, leaving only the diagonals. Because $L_{M,M}$ needs to have positive diagonal entries, $L_{M,M} L_{M,M}^T$ must be positive definite, thus $L_{M,M}$ is the Cholesky factor of A. To derive $L_{M,M}$, a typical approach is to proceed with the Cholesky–Banachiewicz and Cholesky–Crout algorithm on $A$, where entries for the lower triangular factor $L$ are
\begin{equation}
\begin{aligned}
L_{j,j} &= \sqrt{A{j,j} - \sum_{k=1}^{j-1}L_{j,k}L_{j,k}^T} \\
L_{i,j} &= \frac{1}{L_{j,j}}(A_{i,j} - \sum_{k=1}^{j-1}L_{i,k}L_{j,k}^T) \text{ for } i > j \label{eq:crout}
\end{aligned}
\end{equation}
However, for the diagonal entries $L_{j,j}$, instead of using Eq.\ref{eq:crout}, we substitute it with an exponential term $exp(0.5 * O + 2)$ to ensure it is positive, where $O$ is the corresponding unconstrained parameter estimates. Here, $0.5$ was chosen to mimic the square root in the original formula, and $2$ was added to bound initial values of diagonal entries away from zero, given that the default initial values are drawn uniformly from the interval $(-2, 2)$ in \textsf{Stan}. Finally, we update the off-diagonal entries $L_{i,j}$ using the existing formula Eq.\ref{eq:crout}. 
 
In short, in our Bayesian implementation, we set the off-diagonal entries in within-trajectory covariance matrices to be zero, estimate the rest of parameters without constraint using uninformative (non-proper) prior $uniform(-\infty, +\infty)$, substitute the diagonal entries with our exponential term, and finally update the off-diagonal entries. In this way, we are able to estimate the covariance matrix efficiently and guarantee it to be positive semi-definite with our desired constrained form. Once we obtained the covariance matrix, we can then decompose it into correlation matrix $R_\alpha$ and standard deviations in order to estimate temporal associations.

\subsubsection{Estimating inter-block association}
Apart from simultaneously modeling multivariate longitudinal measurements, we want to estimate the association among measurements of interest via the correlations among the FPC scores, where the correlation matrix $R_\alpha$ obtained earlier will play a crucial role. We propose a measure of inter-trajectory association by calculating the mutual information of FPC scores from different measurements. 

We define the inter-trajectory association between measurements $p_1$ and $p_2$ as the mutual information of FPC scores $\boldsymbol{\alpha}_{ip_1}$ and $\boldsymbol{\alpha}_{ip_2}$, $1 \leq p_1, p_2  \leq P$ , with

\begin{align*}
MI(\boldsymbol{\alpha}_{ip_1}, \boldsymbol{\alpha}_{ip_2}) = H(\boldsymbol{\alpha}_{ip_1}) + H(\boldsymbol{\alpha}_{ip_2}) - H(\boldsymbol{\alpha}_{ip_1}, \boldsymbol{\alpha}_{ip_2}), 
\end{align*}

where $H(X)$ is the entropy of $X$ and $H(X) = -E[log(f_{X}(X))]$ with $f_{X}(X)$ being the probability density function of $X$\citep{cover1999elements}.

If $K$-dimensional random variable $X$ follows multivariate normal distribution with covariance matrix $\Sigma$, then according to \citet{ahmed1989entropy}

\begin{align*}
H(X) = \frac{k}{2} + \frac{k}{2} log(2 \pi) + \frac{1}{2} |\Sigma|. 
\end{align*} 

Since the $K$-dimensional FPC scores $\boldsymbol\alpha_i  \sim N(0,\Sigma_\alpha)$, and any subvector of $\boldsymbol\alpha_i$ is of the same structure with the correlation matrix being a submatrix of $R_\alpha$, then according to \citet{arellano2013shannon}, the mutual information of $\boldsymbol{\alpha}_{ip_1}$ and $\boldsymbol{\alpha}_{ip_2}$ could be simplified as

\begin{equation}
\begin{aligned}
MI(\boldsymbol{\alpha}_{ip_1}, \boldsymbol{\alpha}_{ip_2}) =  -\frac{1}{2} log|R_{\alpha\{p_1, p_2\}}|, \label{eq:MI}
\end{aligned}
\end{equation}

where 
\begin{equation*}
R_{\alpha\{p_1, p_2\}} = 
\begin{vmatrix} 
I_{p1} & R_{p_1 p_2} \\
R_{p_1 p_2}^T& I_{p2} \\
\end{vmatrix}.
\end{equation*}

Moreover, we can estimate the conditional inter-trajectory association between any two measurements of interest given the other measurements in the model by calculating the partial mutual information of FPC scores. The conditional inter-trajectory association between measurements $p_1$ and $p_2$ is defined as the partial mutual information of $\boldsymbol{\alpha}_{ip_1}$ and $\boldsymbol{\alpha}_{ip_2}$, $1 \leq p_1, p_2  \leq P$, with

\begin{equation}
\begin{aligned}
MI(\boldsymbol{\alpha}_{ip_1}, \boldsymbol{\alpha}_{ip_2} | \boldsymbol{\alpha}_{i\{1, ..., P \setminus p_1, p_2 \}})
&= H(\boldsymbol{\alpha}_{ip_1}, \boldsymbol{\alpha}_{i\{1, ..., P \setminus p_1, p_2 \}}) +  H(\boldsymbol{\alpha}_{ip_2}, \boldsymbol{\alpha}_{i\{1, ..., P \setminus p_1, p_2 \}}) \\
& \qquad- H(\boldsymbol{\alpha}_{i\{1, ..., P \setminus p_1, p_2 \}}) - H(\boldsymbol{\alpha}_{ip_1}, \boldsymbol{\alpha}_{ip_2} , \boldsymbol{\alpha}_{i\{1, ..., P \setminus p_1, p_2 \}}) \\
&= H( \boldsymbol{\alpha}_{i\{1,...,P \setminus p_2\}}) +  H(\boldsymbol{\alpha}_{i\{1,...,P \setminus p_1\}}) - H(\boldsymbol{\alpha}_{i\{1, ..., P \setminus p_1, p_2 \}}) - H(\boldsymbol{\alpha_i}) \\
&= \frac{1}{2} log|R_{\alpha \{1, ..., P \setminus p_2\}}| + \frac{1}{2} log|R_{\alpha \{1, ..., P \setminus p_1\}}| \\
& \qquad - \frac{1}{2} log|R_{\alpha \{1, ..., P \setminus p_1, p_2\}}| - \frac{1}{2} log|R_{\alpha }|,
\end{aligned}
\end{equation}

where $R_{\alpha \{1, ..., P \setminus p_2\}}$, $R_{\alpha \{1, ..., P \setminus p_1\}}$, and $R_{\alpha \{1, ..., P \setminus p_1, p_2\}}$ are defined in the similar way as $R_{\alpha \{p_1, p_2\}}$ in Eq.(\ref{eq:MI}). 

Inter-trajectory association obtained from this way ranges from 0 to infinity. By analogy with the way Person’s contingency coefficient was obtained, we can apply a simple transformation proposed by \citet{joe1989relative} to obtain a normalized version of the mutual information as

\begin{equation}
MI^*(\boldsymbol{\alpha}_{ip_1}, \boldsymbol{\alpha}_{ip_2}) \coloneqq \sqrt{1-exp[-2MI(\boldsymbol{\alpha}_{ip_1}, \boldsymbol{\alpha}_{ip_2})]}.
\end{equation}

In this way, the inter-trajectory and conditional associations now take its value in [0, 1]. The interpretation is that the closer $MI^*(\boldsymbol{\alpha}_{ip_1}, \boldsymbol{\alpha}_{ip_2})$ or $MI^*(\boldsymbol{\alpha}_{ip_1}, \boldsymbol{\alpha}_{ip_2} | \boldsymbol{\alpha}_{i\{1, ..., P \setminus p_1, p_2 \}})$ is to 1, the higher the temporal association between measurements is. 

\section{Simulation studies}
To evaluate the performance of mSFPCA in modeling multiple temporal measurements, especially in its covariance estimation and temporal association inference, we simulated sparse longitudinal trajectories with three temporal measurements under four different covariance structures. To better mimic the reality, our data was simulated based on an mSFPCA model using parameters initially estimated from a real longitudinal microbiome multi-omics dataset \citep{kostic2015dynamics} in the following way: 

\begin{enumerate}
    \item Applying mSFPCA to model three temporal measurements in the real multi-omics dataset.
    \item Selecting the optimal number of PCs and dimension of basis using PSIS-LOO: the chosen model has the number of PCs as 2, 2, 1, and the number of basis as 6, 5, 5 for each measurement respectively.
    \item Extracting the estimated values for population mean curve $(\boldsymbol\theta_\mu)$, FPC loadings $(\boldsymbol\Theta)$, and residual variance $\sigma_{\boldsymbol\epsilon}$.
\end{enumerate}
	
Then under four distinct covariance structures on FPC scores ($\Sigma_\alpha$), we simulate the trajectories for 100 subjects with an average of 20\% missing data over observations at 10 time points. Observations were randomly deleted to create increasingly sparse functional datasets. In the 1st covariance structure, all PCs are independent; in the 2nd covariance structure, only 1 strong correlation of 0.75 exists across all PCs; in the 3rd covariance structure, 1 strong and 1 medium correlation exists, at values of 0.75 and 0.5 respectively; in the 4th covariance structure, 1 strong, 1 medium and 1 weak correlations exists at strength of 0.75, 0.5 and 0.25. In short, there are increasing dependence structures among PCs as the covariance structure moves from the first to the last. Based on these pre-specified covariance structures and initially estimated parameters, we simulate the sparse longitudinal trajectories as follows:

\begin{enumerate}
    \item Choosing the total number of subjects to be 100, and the number of time points to be 10 in order to place possible time points between $[0,1]$.
    \item  Simulating the observed number of time points for each individual with $n_i \sim Poisson (8)$, where 8 represents the average number  of time points
    across all subjects, and then randomly placing the observed time points in the possible time locations (chosen in the previous step).
    \item  Generating the cubic spline basis matrix $\boldsymbol B_i$ for each subject (orthonormality obtained through Gram-Schmidt orthonormalization).
    \item  Simulating for each subject FPC scores $\boldsymbol{\alpha_i} \sim N(0, \Sigma_\alpha)$ and noise $\boldsymbol\epsilon_i \sim N (0, \sigma_{\boldsymbol\epsilon}^2 I)$.
    \item Obtaining the temporal trajectory for each individual with $\boldsymbol{Y_i} = \boldsymbol{B_i \theta_\mu} + \boldsymbol{B_i \Theta \alpha_i} + \boldsymbol{\epsilon_i}$.
    \item Repeating step 1-- 5 1000 times for each covariance structure, thus generating 4000 simulated datasets in total.
\end{enumerate} 

To evaluate the mSFPCA model performance in simulated data, we want to examine three main results: 1. how well mSFPCA can capture the temporal patterns embodied in the overall mean curve and FPC curves for each temporal measurement; 2. the accuracy of covariance estimation; 3. the inference on temporal associations based on mutual information estimation. 

\begin{figure}
\includegraphics[width=5.8in]{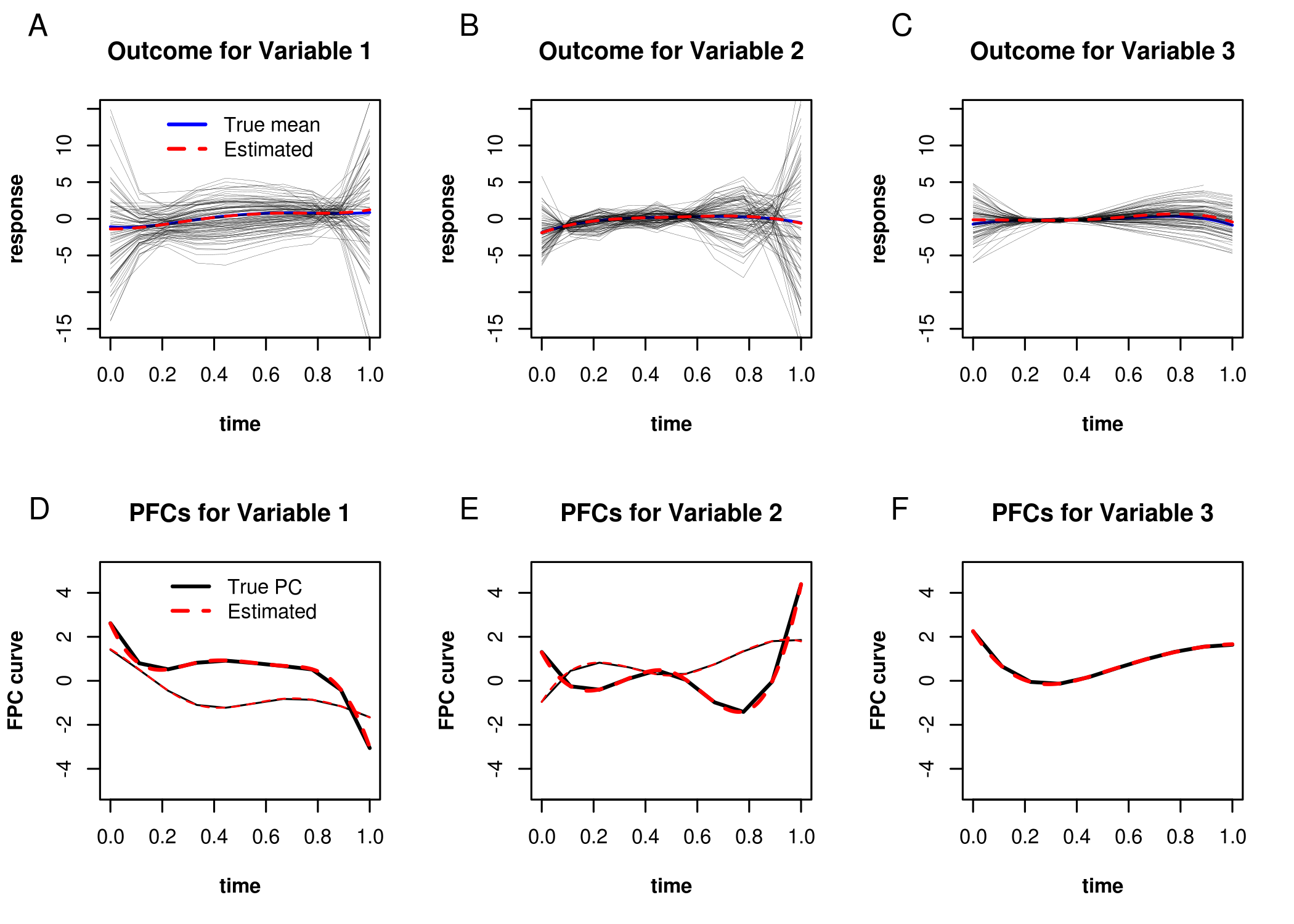}
\caption{Estimated mean and FPC curves from mSFPCA on simulated data with covariance structure I. Estimated (red) vs. true (blue) overall mean curve on simulated trajectories (black) on outcome variable 1 (A), outcome variable 2 (B), and outcome variable 1 (C). Estimated (red) vs. true (black) PC curves on simulated data for outcome variable 1 (D), outcome variable 2 (E), and outcome variable 3 (F).}
\label{sim_curves}
\end{figure}

\begin{figure}
\includegraphics[width=5in]{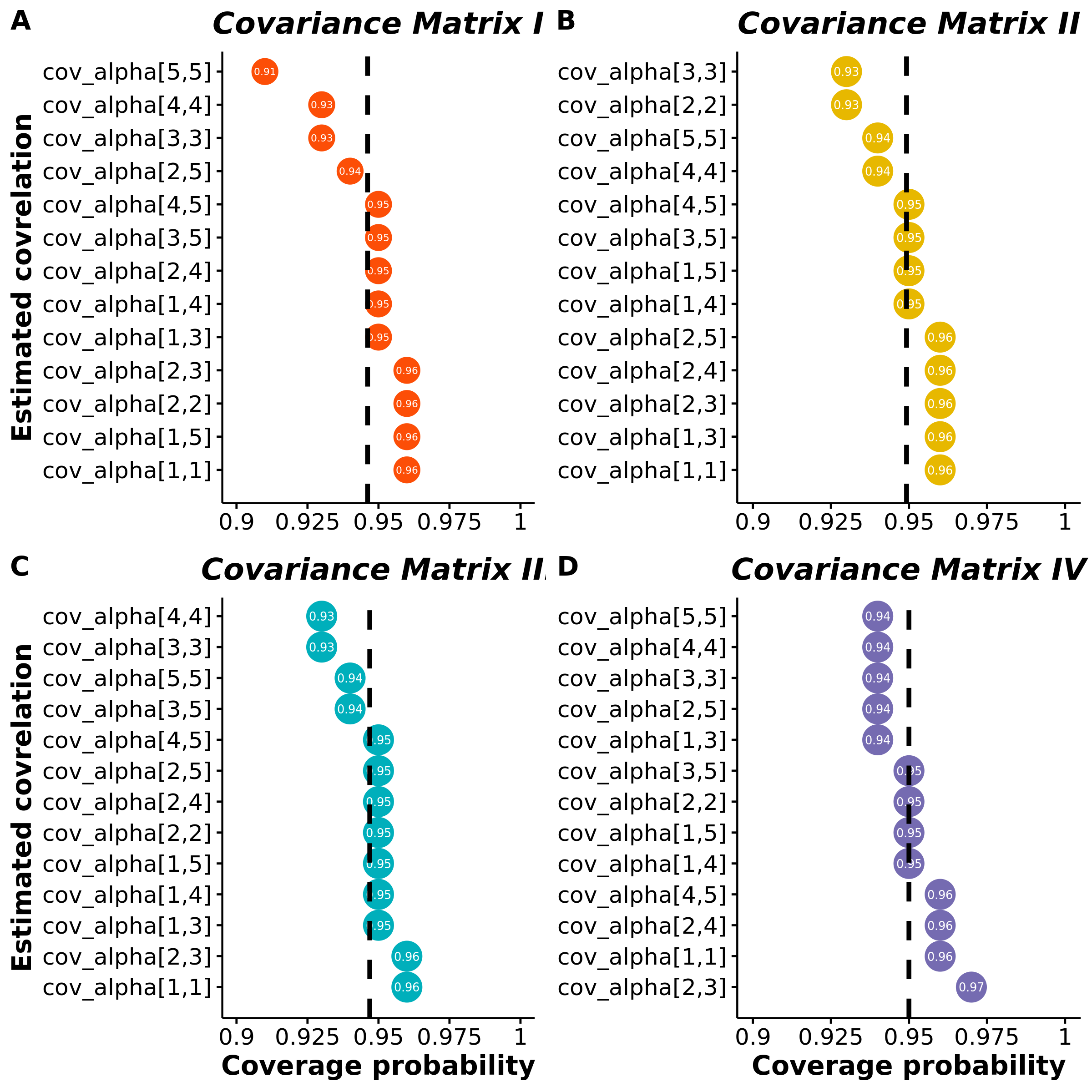}
\caption{Coverage probability of 95\% credible interval on estimated covariance parameters in covariance matrix I (A), II (B), III (C) and IV (D). Values within each dot represent the coverage probability for each estimated covariance parameter, and only values for unique covariance element are displayed. Black dashed lines indicate the average coverage probability across all estimated parameters within each covariance matrix, which is around 95\% in all four simulation scenarios.}
\label{sim_cov_prob}
\end{figure}

Figure~\ref{sim_curves} shows that the estimated overall mean curves and PC curves accurately recovered the ground truth for all three outcome variables under covariance structure I. This accurate capturing of major temporal patterns was seen in other three covariance structures as well (Supplementary Figure 1-3). Figure~\ref{sim_cov_prob} summarizes the performance of covariance estimation across all 4 scenarios in terms of the coverage probabilities of 95\% credible intervals on estimated covariance parameters. The coverage probabilities are lowest in the 1st covariance structure (independent, Figure~\ref{sim_cov_prob}A), improved when more dependence structures are introduced (Figure~\ref{sim_cov_prob}B-D), and reach highest with the 4th covariance structure (having most correlations across PCs, Figure~\ref{sim_cov_prob}D). Despite these subtle differences in the coverage probabilities for each covariance parameter, the average coverage probability across all estimated parameters, represented by the dashed line, is around 95\% within each covariance structure. This indicates that our mSFPCA model is able to estimate the covariance matrix properly, and its performance is affected by the structure of covariance matrix itself: the more sparse the covariance is, the more challenging the estimation. But even with the most sparse scenario (Figure~\ref{sim_cov_prob} A), our mSFPCA model is still able to achieve about 95\% average coverage probability. 

Regarding the inference on temporal associations, Table~\ref{table_sim_MI} shows the mutual information estimates in each simulation scenario, which estimates the temporal association between each pair of temporal measurements. $MI_{ij}$ denotes the normalized mutual information between $i$th and $j$th temporal measurements. When the true MI is zero, the coverage probability is denoted as $0^*$, as the estimated mutual information is always non-negative and thus by construction the posterior distribution will not encompass zero. In the rest of the table, except for the slightly lower coverage probability with 0.92 for $MI_{12}$ in the 3rd scenario, or 0.93 for $MI_{23}$ in the 4th scenario, all the coverage probabilities are close to 95\%. Table~\ref{table_sim_CMI} shows the conditional mutual information estimates in each simulation scenario, which estimates the temporal association between each pair of temporal measurements given the other measurement in the model. $CMI_{ij}$ denotes the normalized conditional mutual information between $i$th and $j$th temporal measurements. All coverage probabilities of non-zero CMIs are close to 95\% on the estimation of conditional mutual information. 

In short, our simulation results have demonstrated the good performance of mSFPCA in modeling sparse longitudinal data with multiple temporal measurements and providing valid inference on temporal associations.

\begin{table}
\caption{Mutual information estimates for each simulation scenario}
\label{table_sim_MI}
\begin{tabular}{@{}lcrcrrr@{}}
\hline
&& & &\multicolumn{3}{c}{95\% credible interval} \\
\cline{5-7}
Simulation scenario &Parameter &
\multicolumn{1}{c}{Truth} &
\multicolumn{1}{c}{Median} &
\multicolumn{1}{c}{Cov.prob.} &
\multicolumn{1}{c}{2.5\%} &
\multicolumn{1}{c@{}}{97.5\%} \\
\hline
{Covariance I} & $MI_{12}$ & 0 & 0.26 & 0* & 0.11 & 0.42 \\
          & $MI_{13}$  & 0   & 0.17 & 0* & 0.04 & 0.34 \\
          & $MI_{23}$   & 0   & 0.18 & 0*  & 0.04 & 0.35 \\[6pt]
{Covariance II} & $MI_{12}$   & 0  & 0.26 & 0* & 0.11 & 0.42  \\
          & $MI_{13}$   & 0   & 0.17 & 0* & 0.04 & 0.34  \\
          & $MI_{23}$   & 0.75  & 0.75 & 0.96 &
          0.65  & 0.83  \\[6pt]
{Covariance III} & $MI_{12}$   & 0.5  & 0.54 & 0.92 & 0.39 & 0.66\\
          & $MI_{13}$   & 0   & 0.17 & 0*   & 0.04 & 0.34  \\
          & $MI_{23}$   & 0.75  & 0.75 & 0.96 
          & 0.65 & 0.83 \\[6pt]          
{Covariance IV} & $MI_{12}$   & 0.5 & 0.54 & 0.94 & 0.38 & 0.66 \\
          & $MI_{13}$  & 0.25  & 0.29 & 0.94   & 0.12 & 0.46  \\
          & $MI_{23}$   & 0.75  & 0.75 & 0.93  & 0.66 & 0.83  \\
\hline
\end{tabular}
\end{table}

\begin{table}
\caption{Conditional mutual information estimates for each simulation scenario}
\label{table_sim_CMI}
\begin{tabular}{@{}lcrcrrr@{}}
\hline
&& & &\multicolumn{3}{c}{95\% credible interval} \\
\cline{5-7}
Simulation scenario &Parameter &
\multicolumn{1}{c}{Truth} &
\multicolumn{1}{c}{Median} &
\multicolumn{1}{c}{Cov.prob.} &
\multicolumn{1}{c}{2.5\%} &
\multicolumn{1}{c@{}}{97.5\%} \\
\hline
{Covariance I} & $CMI_{12}$ & 0 & 0.26 & 0* & 0.11 & 0.42 \\
          & $CMI_{13}$  & 0   & 0.17 & 0* & 0.04 & 0.34 \\
          & $CMI_{23}$   & 0   & 0.18 & 0*  & 0.04 & 0.35 \\[6pt]
{Covariance II} & $CMI_{12}$   & 0  & 0.26 & 0* & 0.12 & 0.42  \\
          & $CMI_{13}$   & 0   & 0.17 & 0* & 0.04 & 0.34  \\
          & $CMI_{23}$   & 0.75  & 0.75 & 0.95 & 0.65  & 0.83  \\[6pt]
{Covariance III} & $CMI_{12}$   & 0.76  & 0.77 & 0.94 & 0.68 & 0.84\\
          & $CMI_{13}$   & 0.66   & 0.66 & 0.95   & 0.53 & 0.76  \\
          & $CMI_{23}$   & 0.87  & 0.87 & 0.96 
          & 0.81 & 0.91 \\[6pt]          
{Covariance IV} & $CMI_{12}$   & 0.81 & 0.82 & 0.95 & 0.74 & 0.88 \\
          & $CMI_{13}$  & 0.76  & 0.76 & 0.95   & 0.66 & 0.83  \\
          & $CMI_{23}$   & 0.89  & 0.90 & 0.95  & 0.85 & 0.93  \\
\hline
\end{tabular}
\end{table}

\section{Real data application}
For the real data application, we want to model multiple temporal measurements simultaneously in a large and challenging dataset, with a special interest in utilizing conditional mutual information to infer temporal association. This dataset comes from the type 2 diabetes (T2D) longitudinal studies in the Integrative Human Microbiome Project \citep{integrative2014integrative}. In this example, an over 3 years' study has been conducted in approximately 100 individuals at high risk for T2D, in order to better understand the biological changes that occur during the onset and progression of T2D. Multiple sample types were collected from the study participants every 2-3 months during their healthy periods, with more frequent sampling during periods of respiratory illness and other environmental stressors. These data include multi-omics assays such as stool microbiome data using 16S rRNA sequencing, host protein expression profiles in fecal samples using LC-MS/MS, and cytokine profiles that quantify the levels of 50 diverse inflammatory proteins and insulin peptides in host serum, as well as standard clinical tests results like hemoglobin A1c (HbA1c), insulin and glucose. Moreover, behavior changes of patients, such as emotional and psychological stress, were documented using the Perceived Stress Scale instrument. Our outcomes of interest are the longitudinal pattern of Shannon diversities in bacteria, proteins and cytokines, and of clinical test results on HbA1c. Shannon diversity is defined as $Shannon = - \sum_{i=1}^{S} p_i ln(p_i) $, where $S$ is the total number of species, and $p_i$ is the relative proportion of species $i$ relative to the entire population. Shannon diversities in bacteria, proteins and cytokines are chosen over specific features in our application, because they have higher predictive power of individuals' diabetic status than specific bacteria, proteins or cytokines. Hence, we are particularly interested in utilizing mutual information to investigate which omics data (based on Shannon diversity) have strongest association with HbA1c, and whether additional omics data improve the temporal association based on conditional mutual information. 

\begin{figure}
\includegraphics[width=5.5in]{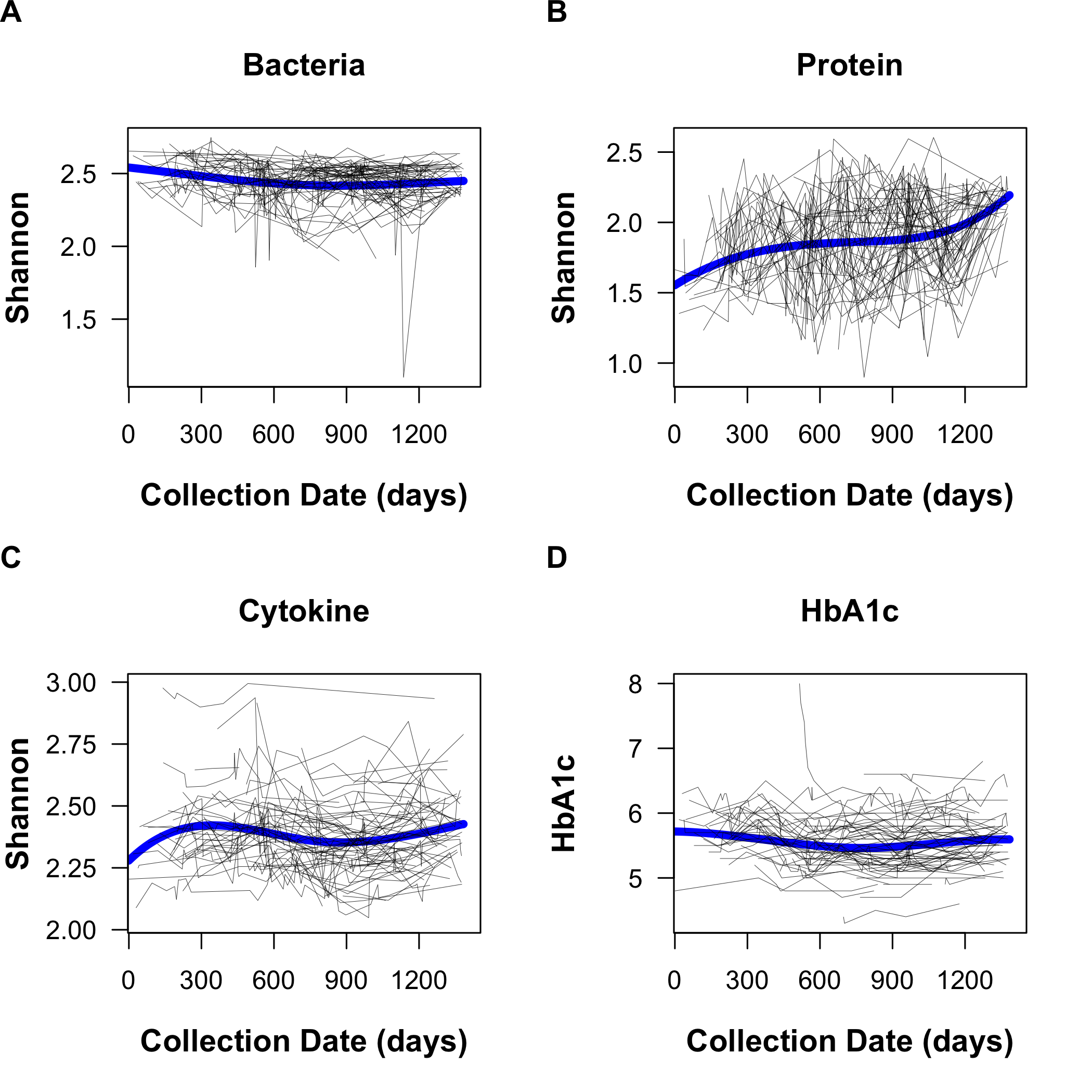}
\caption{Estimated mean curves from mSFPCA application on type 2 diabetes multi-omics dataset. (A) Estimated population mean curve (blue) for Shannon bacterial diversity on observed individual trajectories (black). (B) Estimated population mean curve (blue) for Shannon protein diversity on observed individual trajectories (black). (C) Estimated population mean curve (blue) for Shannon cytokine diversity on observed individual trajectories (black). (D) Estimated population mean curve (blue) for HbA1c on observed individual trajectories (black).}
\label{t2d_fig1}
\end{figure}

\begin{figure}
\includegraphics[width=5in]{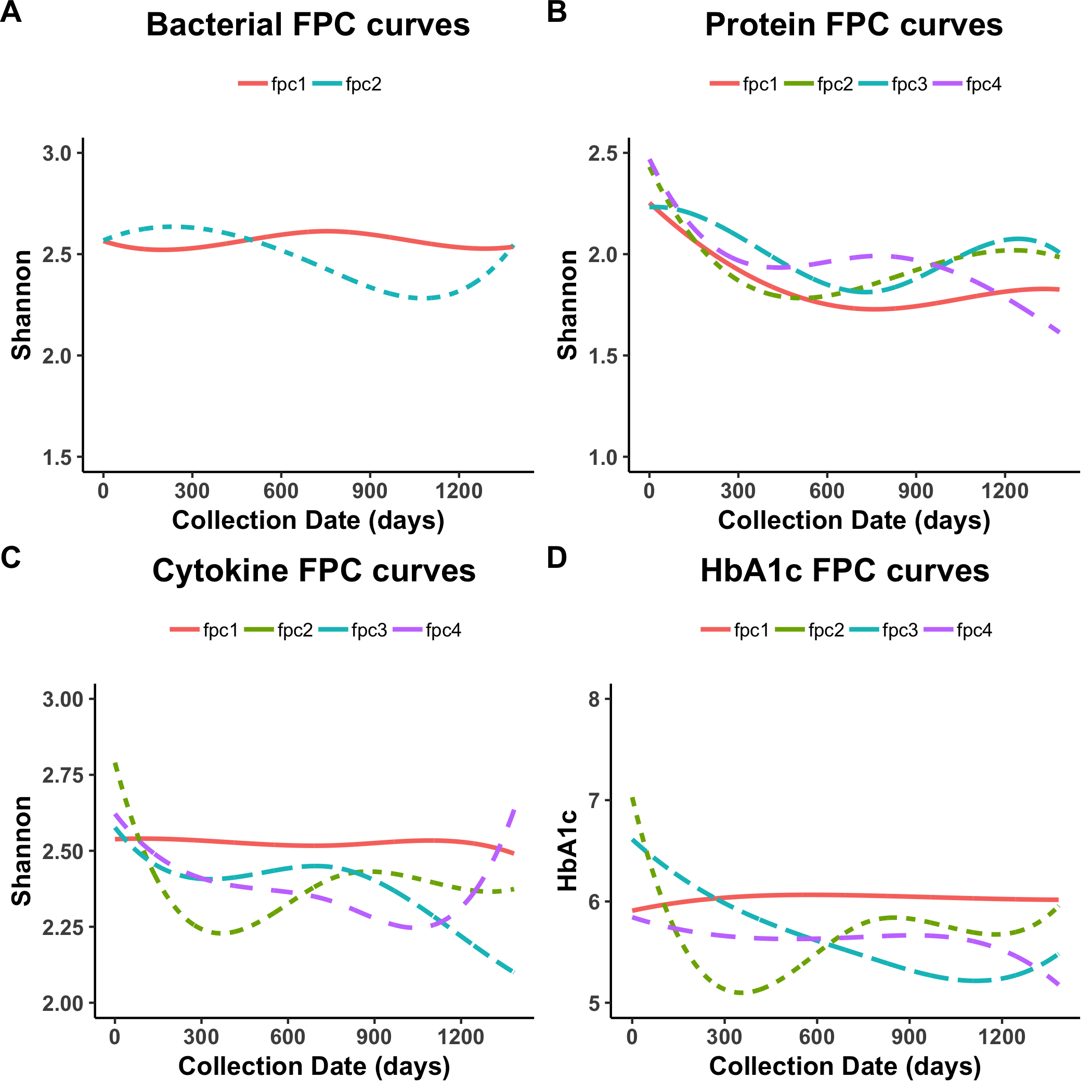}
\caption{Estimated FPC curves from mSFPCA application on type 2 diabetes multi-omics dataset. (A) Trends of variability in Shannon bacterial diversity captured by 2 principal component curves. (B) Trends of variability in Shannon protein diversity captured by 4 principal component curves. (C) Trends of variability in Shannon cytokine diversity captured by 4 principal component curves. (D) Trends of variability in HbA1c captured by 4 principal component curves.}
\label{t2d_fig2}
\end{figure}

The estimated mean curves in Figure~\ref{t2d_fig1} show different temporal trends in each outcome, where Shannon bacterial diversity decreases slowly over time, protein diversity increases steadily over time, cytokine diversity increases over the first 300 days, decreases between day 300 and 900, and then increases, and HbA1c decreases during the first 2 years, and increases slightly afterwards. As indicated by the observed trajectories for each individual (black curves), there are great subject-level variations in each outcome. This additional temporal information is captured by the FPC curves in Figure~\ref{t2d_fig2}. Figure~\ref{t2d_fig2}A shows the first two PCs in Shannon bacterial diversity, of which PC 1 explains 79\% and captures variation around day 750, while PC 2 explains 21\% of the variation and emphasizes variation around day 300 and 1100. Figure~\ref{t2d_fig2}B shows the first four PCs in Shannon protein diversity, of which the first two PCs explain over 90\% of the variance. The first PC captures variation around day 750, and the second PC emphasizes variation around day 450 and 1200. Figure~\ref{t2d_fig2}C shows the first four PCs in Shannon cytokine diversity, of which PC 1 explains 83\% and exhibits an almost flat curve over time, while PC 2 explains 13\% of the variation and emphasizes variation around day 300 and 800. Figure~\ref{t2d_fig2}D shows the first four PCs in HbA1c: PC1 exhibits a slight increasing curve over time, accounting for 70\% variation, and PC 2 captures variation around day 300 and 800, explaining for an additional 21\% variation. In short, although principal patterns in each measurement vary, changing time points are pretty consistent, suggesting coherent responses to changes in patients’ mental or physical conditions. 

Among omics' temporal associations with standard clinical test result HbA1c, Table~\ref{t2d_table} suggests that Shannon protein diversity has the highest association with HbA1c, at an estimated mutual information of 0.91 with 95\% credible interval (0.786, 0.971). Cytokine diversity is the second highest, with MI at 0.849 (0.668, 0.954), and bacteria diversity has the lowest association at 0.71 (0.441, 0.854). However, when information about other omics measurements are provided, all the pairwise temporal associations increase to over 0.95, as indicated by the conditional MI results. Regarding temporal associations among omics measurement, Shannon protein and bacteria diversities have highest temporal association, with mutual information at 0.982 (0.875, 0.999); Shannon protein and cytokine diversities also have high association at 0.966 (0.886, 0.998); the association between Shannon bacteria and cytokine is medium at 0.798 (0.454, 0.977). Similar to earlier results, when information about other measurements are available, all conditional information increase to 0.99. In short, host protein expression profiles data has highest temporal association with patients’ diabetes status (i.e. HbA1c), but this information can still be further improved with additional omics data. 

\begin{figure}
\includegraphics[width=5in]{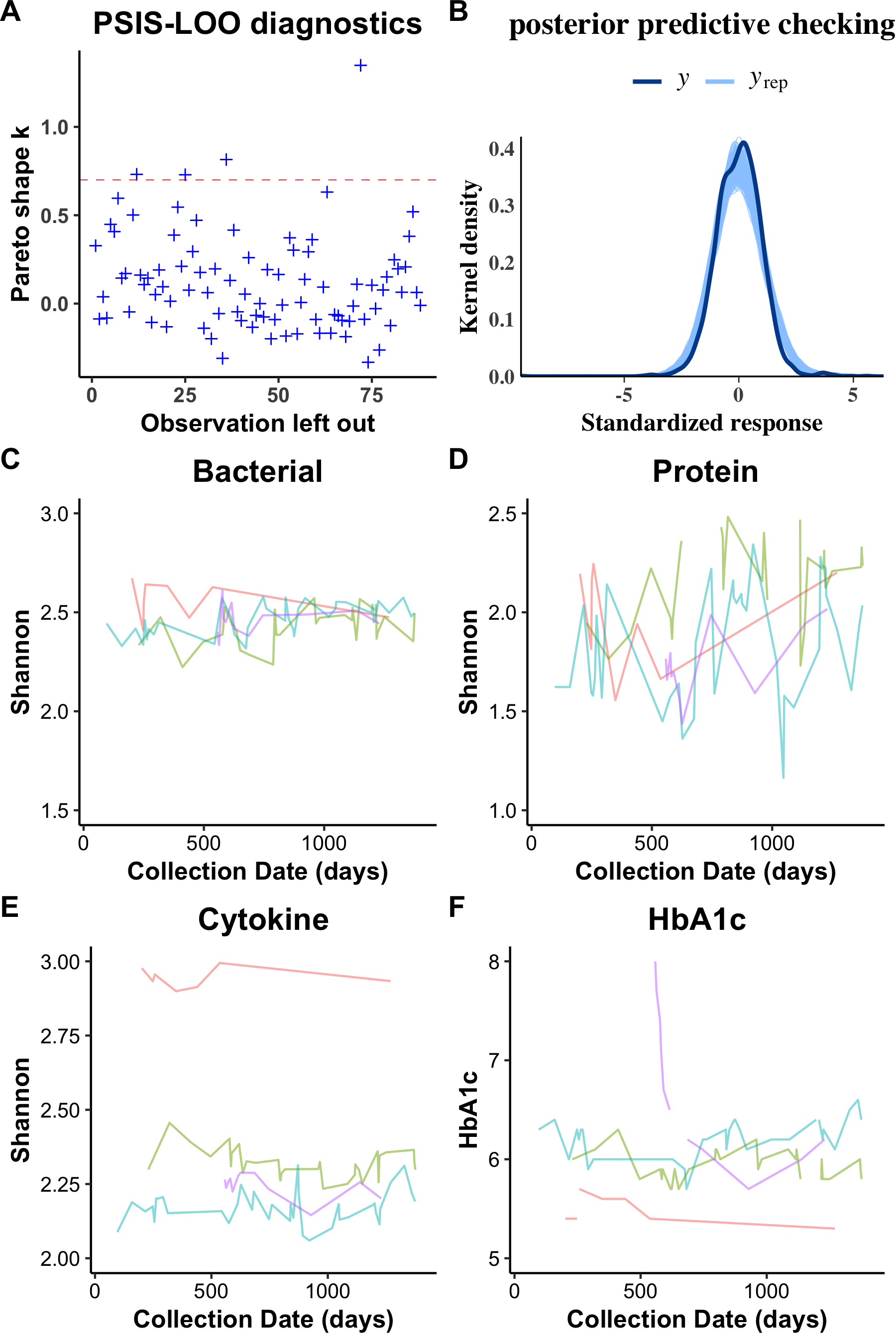}
\caption{Graphical model diagnostics and examination of outliers for mSFPCA application on type 2 diabetes multi-omics dataset. (A) Scatterplot of estimated Pareto shape parameter $\hat k$ in PSIS-LOO diagnostic plot: all $\hat k$’s except 4 are lower than the warning threshold 0.7. (B) Graphical posterior predictive plot: kernel density estimate of the observed dataset $y$ (dark curve), with kernel estimates for 100 simulated dataset $y_{rep}$ drawn from the posterior predictive distribution (thin, lighter lines). (C) Observed Shannon bacterial diversity for 4 outliers detected by PSIS-LOO diagnostic plot in (A). (D) Observed Shannon protein diversity for 4 outliers. (E) Observed Shannon cytokine diversity for 4 outliers. (F) Observed HbA1c for 4 outliers.
}
\label{t2d_fig3}
\end{figure}

We need model diagnostics to conclude on the validity of our mSFPCA application. The optimal model selected by PSIS-LOO has 2 PCs for Shannon bacterial diversity, and 4 PCs for the other measurements, and the number of internal knot is chosen to be one for all outcomes. PSIS-LOO diagnostics in Figure~\ref{t2d_fig3}A show that the selected mSFPCA model fit the majority of the data well, except for 4 outliers with Pareto shape k values higher than the warning threshold 0.7.  
Graphical posterior predictive checks in Figure~\ref{t2d_fig3}B suggests good model fit as the simulated data from the posterior predictive distribution was able to cover the distribution of observed outcomes well. Figure~\ref{t2d_fig3}C-F highlight the observed trajectories of the 4 outliers detected by PSIS-LOO diagnostic plot. The red subject has highest curve in Shannon cytokine diversity and low value in HbA1c. A closer look at his/her metadata shows that this subject went through stages of healthy, infection and back to healthy. The green subject shows high oscillation pattern in Shannon protein  diversity, as he/she oscillated between stages of healthy, inflammation, and infection. The blue subject exhibits high oscillation pattern in Shannon protein diversity, because he/she went through a complicated interweaving stages of healthy, inflammation, infection, post-travel and allergy. The purple subject, who has the highest Pareto shape k value in Figure~\ref{t2d_fig3}A, experienced drastic change in HbA1c, as he/she went through stages of infection, stress and back to healthy. In short, our mSFPCA model generally fits this dataset well, and our diagnostic tools were able to highlight biologically meaningful outliers for further examination. 

\begin{table*}
\caption{Mutual information estimates for type 2 diabetes multi-omics dataset application}
\label{t2d_table}
\begin{tabular}{@{}lrrrrc@{}}
\hline
temporal associations with HbA1c 
& \multicolumn{1}{c}{$MI (95\% CI)$}
& \multicolumn{1}{c}{$CMI (95\% CI)$} \\
\hline
HbA1c---protein    & 0.910 (0.786, 0.971) & 0.994 (0.968, 0.999)  \\
HbA1c---cytokine    & 0.849 (0.668, 0.954)  & 0.986 (0.951, 0.999)  \\
HbA1c---bacteria    & 0.710 (0.441, 0.854)  & 0.957 (0.820, 0.999)  \\
\hline
temporal associations among omics
& \multicolumn{1}{c}{$MI (95\% CI)$}
& \multicolumn{1}{c}{$CMI (95\% CI)$} \\
\hline
protein---bacteria    & 0.982 (0.875, 0.999) & 0.999 (0.996, 0.999)  \\
protein---cytokine    & 0.966 (0.886, 0.998)  & 0.999 (0.997, 0.999)  \\
bacteria---cytokine    & 0.798 (0.454, 0.977)  & 0.995 (0.958, 0.999)  \\
\hline
\end{tabular}
\end{table*}

\section{Discussion}

Here, we have proposed multivariate sparse functional PCA, an extension to the sparse functional principal components analysis, to modeling multiple trajectories simultaneously. The methodological novelty lies in the computationally efficient Bayesian covariance matrix estimation, where we utilized a Cholesky decomposition to guarantee it to be positive semi-definite under the constrained form of diagonal within-trajectory covariance and arbitrary form of between-trajectory covariance structure. Moreover, we utilized mutual information to assess marginal and conditional temporal associations, providing. Finally, our Bayesian implementation in \textsf{Stan} enables the usage of PSIS-LOO for efficient model selection, and visual model diagnostic methods, such as examining the estimated shape parameters from PSIS-LOO and utilizing the graphical posterior predictive checks, to evaluate the validity of mSFPCA models and highlight potential outliers. 

In both our real-data based simulations and application to longitudinal microbiome multi-omics datasets, we have demonstrated that mSFPCA is able to accurately uncover the underlying principal modes of variation over time, including both the average population pattern and subject-level variation, and estimate the temporal associations properly. These enabled us to detect biologically meaningful signals in a large and challenging longitudinal cohort with irregular sampling, missing data, and four temporal measurements. Moreover, the model diagnostics plots from real data application show that mSFPCA can provide reliable model fitting to real microbiome multi-omics dataset. All these results highlight the great value of our method in modeling longitudinal data with multiple temporal measurements. Though we applied mSFPCA to microbiome data in this paper, the method is in fact a general framework that can be applied to a wide range of multivariate longitudinal data. 

One limitation of our method is that we assume the principal component scores and residuals to be normally distributed as in the original SFPCA model. This normality assumption would restrict our method from being applying to highly skewed trajectories, for example However, improper application of the method to such data could be detected by the model diagnostic tools we provide, e.g., the graphical posterior predictive model checks. Users could also modify the mSFPCA model by incorporating alternative prior distributions, for example, a t-distribution with a low degree of freedom to capture heavy tails in the distribution of principal component scores, which can be easily implemented in \textsf{Stan}. Finally, since the mSFPCA model is implemented in \textsf{Stan}, a programming language with a very active user base, this method will be able to be updated with more efficient MCMC sampling algorithms and also incorporate other groundbreaking model selection and diagnostic techniques whenever they become available. Hence, we believe that the mSFPCA method will become a useful and up-to-date tool for researchers in various fields to analyze longitudinal data with multiple measurements in order to detect complex temporal associations.

\section{Acknowledgement}
RK was supported by NIH under grant 1DP1AT010885, NIDDK under grant 1P30DK120515, and CCFA under grant 675191. WT was supported by NIH/NIMH under grants MH120025 and MH122688.

%
%

%
%
 
\begin{supplement}
\textbf{Supplement to "multivariate Sparse Functional Principal Components Analysis for Longitudinal Microbiome Multi-Omics Data"}. Three supplementary figures for simulation results are included in the supplementary materials.
\end{supplement}


\bibliographystyle{imsart-nameyear} 
\bibliography{bibliography}       


\end{document}


\begin{frontmatter}
\title{Multi-Block Sparse Functional Principal Components Analysis for Longitudinal Microbiome Multi-Omics Data}
\runtitle{Multi-Block Sparse Functional Principal Components Analysis for Longitudinal Microbiome Multi-Omics Data}

\begin{aug}
\author[A]{\fnms{Lingjing} \snm{Jiang}\ead[label=e1,mark]{lij014@health.ucsd.edu}},
\author[B]{\fnms{Chris} \snm{Elord}\ead[label=e2]{chris.elrod@juliacomputing.com}},
\author[C]{\fnms{Jane} \snm{J. Kim}\ead[label=e6]{janekim@health.ucsd.edu}},
\author[D]{\fnms{Austin}
\snm{D. Swafford}\ead[label=e3]{adswafford@eng.ucsd.edu}},
\author[C,D,E,F]{\fnms{Rob} \snm{Knight}\ead[label=e4]{robknight@health.ucsd.edu}}
\and
\author[A]{\fnms{Wesley} \snm{K. Thompson}\ead[label=e5, mark]{wkthompson@health.ucsd.edu}}
\address[A]{Herbert Wertheim School of Public Health and Human Longevity Science, 
University of California San Diego, 
\printead{e1,e5}}

\address[B]{Julia Computing, 
\printead{e2}}

\address[C]{Department of Pediatrics,
University of California San Diego, 
\printead{e6,e4}}

\address[D]{Center for Microbiome Innovation,
University of California San Diego,
\printead{e3,e4}}

\address[E]{Department of Computer Science and Engineering,
University of California San Diego, 
\printead{e4}}

\address[F]{Department of Bioengineering,
University of California San Diego,
 \printead{e4}}

\end{aug}
\end{frontmatter}

\beginsupplement
\section*{Supplementary Materials}
Three supplementary figures for simulation results are included as supplementary materials.

\begin{figure}[h]
 \centerline{\includegraphics[width=5in]{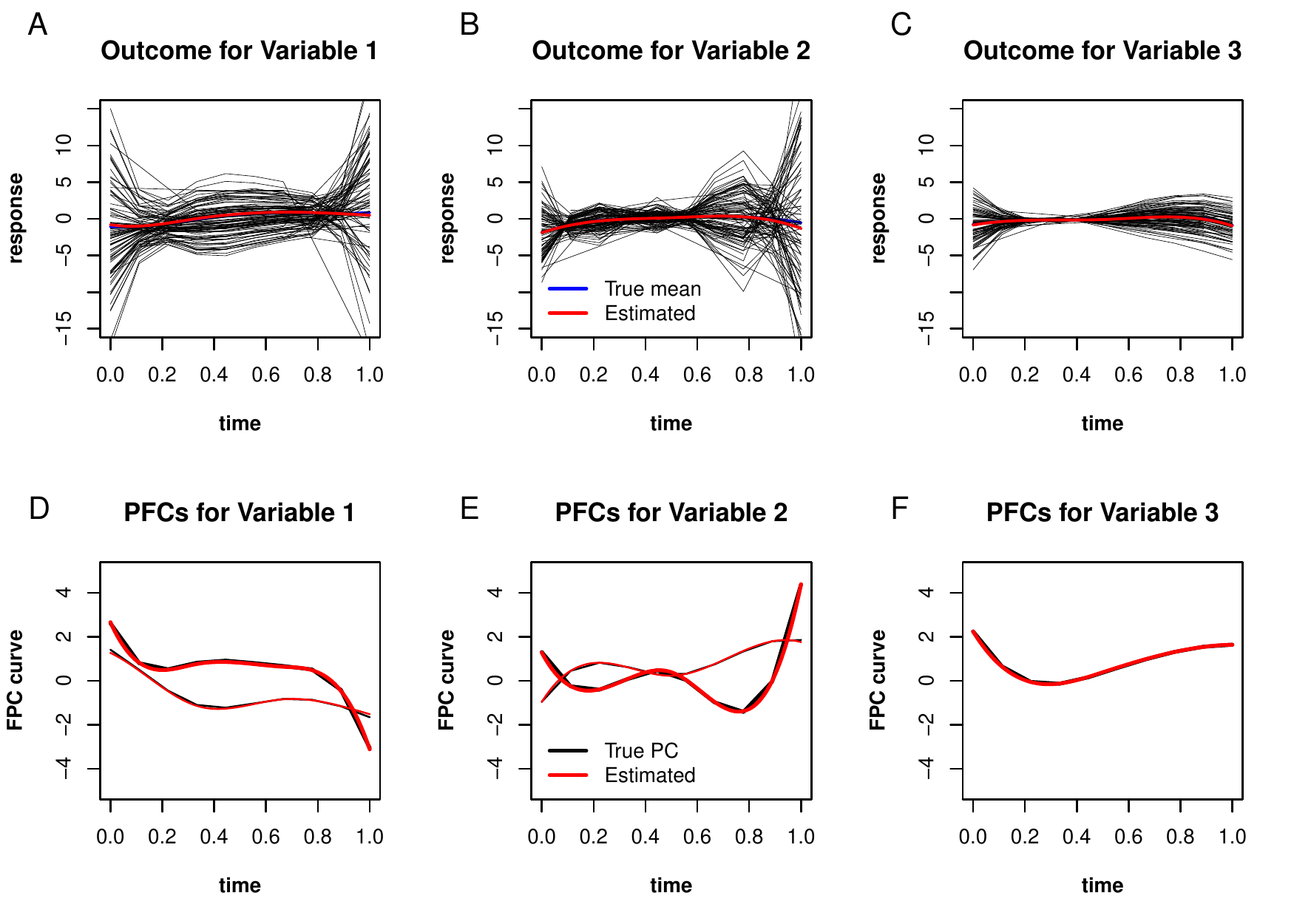}}
\caption{Estimated mean and FPC curves from mSMFPCA on simulated data with covariance structure II. Estimated (red) vs. true (blue) overall mean curve on simulated trajectories (black) on outcome variable 1 (A), outcome variable 2 (B), and outcome variable 1 (C). Estimated (red) vs. true (black) PC curves on simulated data for outcome variable 1 (D), outcome variable 2 (E), and outcome variable 3 (F).}
\label{f:figS1}
\end{figure}

\begin{figure}[h]
 \centerline{\includegraphics[width=5in]{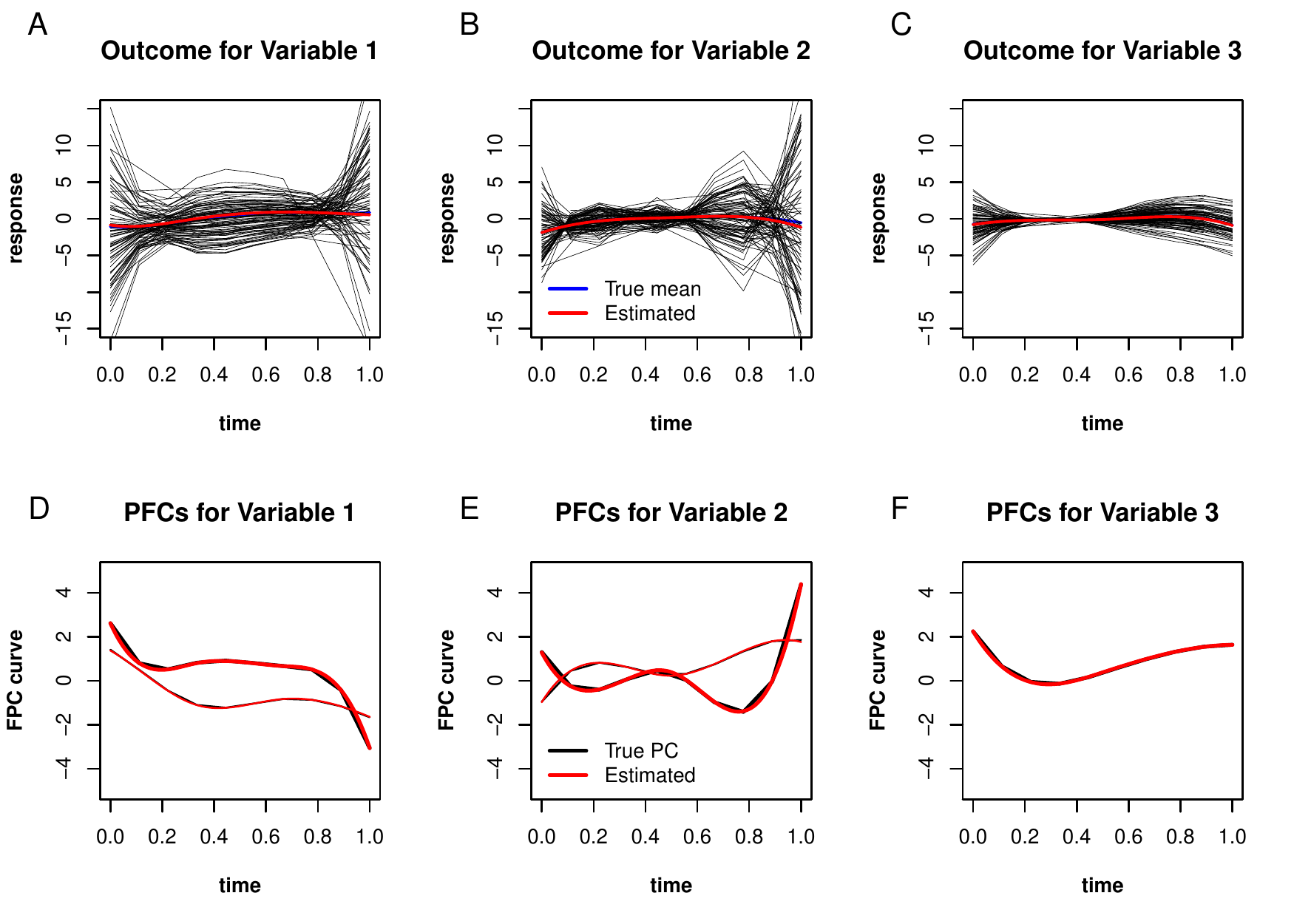}}
\caption{Estimated mean and FPC curves from mSMFPCA on simulated data with covariance structure III. Estimated (red) vs. true (blue) overall mean curve on simulated trajectories (black) on outcome variable 1 (A), outcome variable 2 (B), and outcome variable 1 (C). Estimated (red) vs. true (black) PC curves on simulated data for outcome variable 1 (D), outcome variable 2 (E), and outcome variable 3 (F).}
\label{f:figS2}
\end{figure}

\begin{figure}[h]
 \centerline{\includegraphics[width=5in]{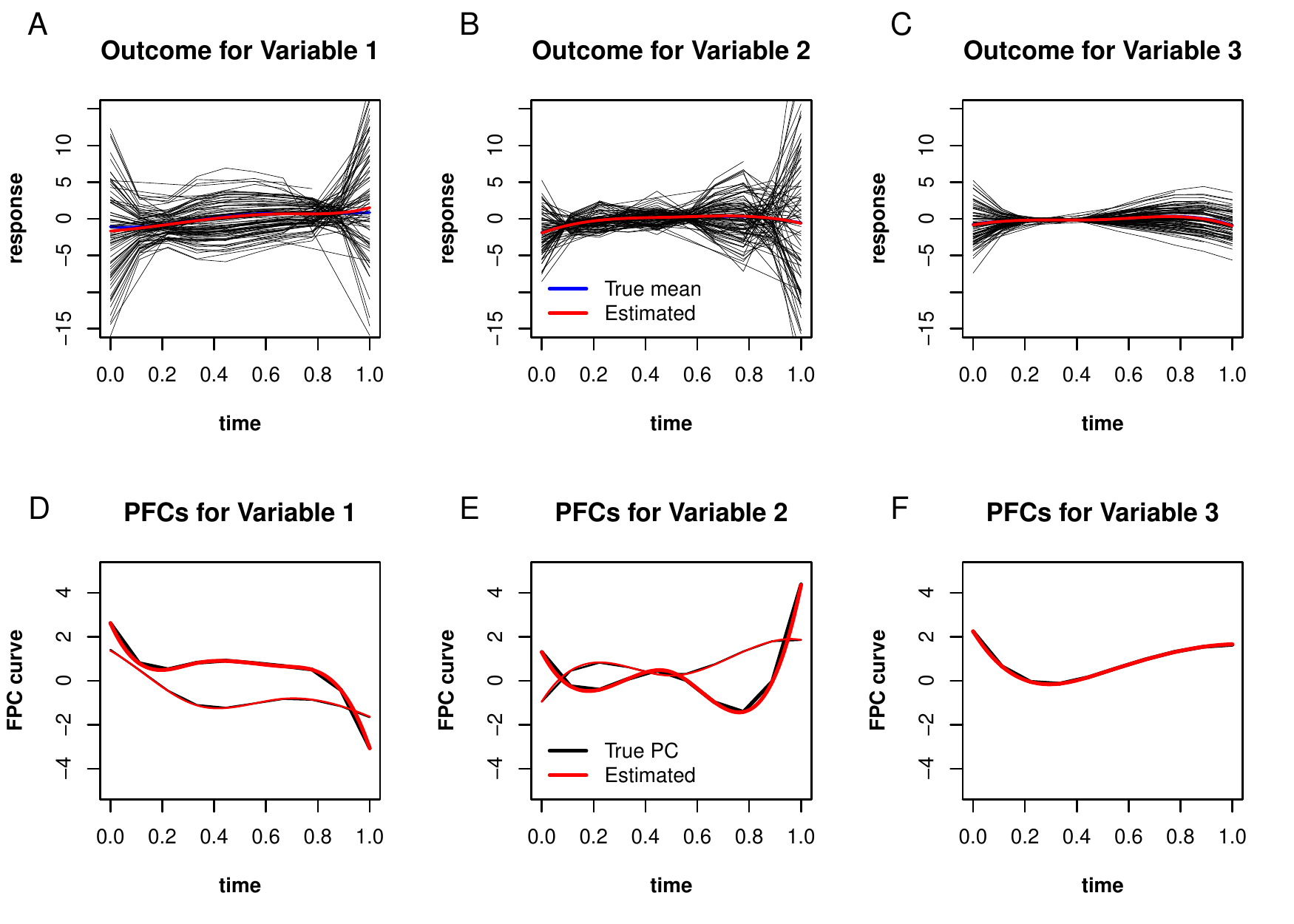}}
\caption{Estimated mean and FPC curves from mSMFPCA on simulated data with covariance structure IV. Estimated (red) vs. true (blue) overall mean curve on simulated trajectories (black) on outcome variable 1 (A), outcome variable 2 (B), and outcome variable 1 (C). Estimated (red) vs. true (black) PC curves on simulated data for outcome variable 1 (D), outcome variable 2 (E), and outcome variable 3 (F).}
\label{f:figS3}
\end{figure}